\documentstyle[sao2,psfig]{article}
\newcommand{\g}{$\gamma$}
\issue{1997, 43, 5-17}
\begin{document}
 \title{Deep  photometric study of the  region of the $\gamma$--ray burst
localization of June 13, 1979.} \author{Sokolov V.V.\inst{a} \and
Kurt V.G.\inst{b} \and Zharikov S.V.\inst{a} \and Kopylov A.
I.\inst{a} \and Berezin A.V.\inst{b}}
\institute{\saoname
 \and  Astro Space Center of Russian Acad. of Sciences,
117810, Moscow, Russia}
\date{April 17, 1997}{May 12, 1997}
\maketitle \begin{abstract} In the  framework  of
the hypothesis  on potential sources  of some \g--ray bursts
being single compact objects  of  neutron star type with finite surface
temperature in the  vicinity  of the Sun a  method  of   search  in the
optical range for a corresponding candidate in the              region of a
powerful short burst localization is considered. The localization region of
the GRB\,790613 event is studied in the {\it B, V} and {\it R} filters.  $B-V$
and $V-R$ colors of  196 objects with a $S/N > 4$ are measured up to  the
average practical photometric limit  $B < 25.7, V < 25.7\ {\rm  and}\  R <
25.6$ for the whole obtained mosaic of CCD images. No blue objects with $B-V
< 0$\ {\rm and}\ $V-R <0$ were found up to $B \approx 25.2$, which can be
interpreted as the  absence  of  a  compact object  with a temperature $>
1.5\cdot 10^{5}\,\rm K$ in this  direction up  to  a distance  $\approx40$\,
pc.  A probable common upper limit to the temperature of the supposed
compact objects in the vicinity of the Sun is indicated based on the
accumulated optical  and X--ray  data.  The study of blue ($B-V < 0$) objects
in the smallest (in  the northern sky) \g--ray box GRB790613, in view of the
results of the search for point--like sources in the soft X--ray in the
localization regions  of bright \g--ray bursts, does not exclude  the
existence  of such ``cold'' ($\la 10^{5}\,\rm K$) objects, which  could
give an essential contribution to the observed density of hidden matter close
to the Sun and some of them could be related with the closest  sources  of
\g--ray bursts.  Several interesting objects were also found  in (or near)
the localization region of the GRB\,790613 event.   \end{abstract}

\section{Introduction}
Here we define the strategy of  search for  optical  candidates
for the sources of \g--ray bursts first of all as a  direct  test
of their galactic origin, in particular,  allowing  for  the  fact
that potential sources of some part of the observed  bursts can be
objects of hidden mass in the Galaxy. It is evident that deep  optical
observations with consideration of observations in other  wavelength
ranges must reduce  the  number  of  theoretic  variants  explaining  the
bursts, at least for a chosen class of galactic  models  discussed
in the literature (Boer et al., 1993;  Greiner  et al., 1995;
Hartmann, 1994).

Formulating below the statement of an  observational problem  at  the
6 m telescope, we mean single compact objects of
neutron star type  cooled down to a temperature  $\approx10^{5}\,\rm K$ and
lower.  Analogous  considerations  have already been  used   in   the
literature in  interpretation of the results of deep photometric
study  of  several  \g--ray  boxes  (Hartmann  et al., 1989; Motch  et
  al., 1990).  Ultimately,  in  any galactic model of the sources of
\g--ray bursts, in  which  attempts of allowing for new data have been
made, a problem can be  stated of search for {\it closest} objects in
other wavelength ranges.  In this case we are concerned with
the sources of short \g--ray bursts, that show a $\rm -3/2\,\, logN-logP$
power law at high intensities (Kouveliotou  et al., 1993). In particular,
these can be optical objects of low luminosity related to  these bursts and
having a density  distribution  in the {\it B, V, R} bands close to that of a
black body.  With the help of  deep CCD photometry of the localization region
of the \g--ray  burst GRB\,790613 we will show which distance limitation can
be obtained from  direct  optical  observations.  In so  doing  we  will use
basically only the assumption of compact character of the supposed potential
source of a \g--ray burst and the fact that the number of such objects in the
vicinity of the Sun can not be greater than it is allowed by the estimates of
hidden mass density.  Further we use also the results of surveys and
point--like observations in the soft X--rays by the ROSAT satellite.

Apparently, similar objects, single pulsars Ge\-minga and PSR\,0656+14 with
relatively  low  surface  temperatures  and  probably situated in the
vicinity of the Sun have  already been observed in the optical (Halpern \&
Tytler, 1988; Caraveo et al., 1994) as sources of $25-25\fm5$. By the
 example of one  of the most powerful \g--ray bursts --- the short  event
 GRB 790613 localized with the help of several space detectors with  the
best (in the northern sky) accuracy (Barat et al., 1984), whose localization
error box has already been studied very well in the X--rays (see  Greiner et
al., 1995;  and ROSAT X--ray Images, 1994a)   and in the optical range
(Ricker  et al., 1986; Harrison et al., 1994; Vrba  et al., 1995),  and
supplementing these data with our own deep optical observations of the
 \g--ray box GRB\,790613, we will  show here that an object of
blue  optical  candidate type for Geminga can, in principle, be detected in
study of such  an  IPN \g--box in the three standard filters {\it B,~V,~R}.

In Section 2 the statement of a  problem                 of  the
deep CCD photometric search  for  weak  blue  stellar--like  objects  in
relatively small \g--ray boxes at the  6 m  telescope is  expounded.
The obtained photometric data for the GRB\,790613 event  field  are  adduced
in Section 3.2.  The method of photometric search for such objects in the
\g--ray box GRB\,790613, which are ``black  body''--like  by  {\it BVR},
and the discussion of the results of deep optical  observations  of
this box are presented in Section 3.3.
The estimates of a probable value of proper  motions  per
year for all objects up to $V = 25^m$   are presented  in
Section 3.4.
 Section 3.5 is devoted  to
some other interesting objects found in (or near)  the  region  of
localization of the GRB\,790613 event.
\begin{table}
\caption{ The log--book of observations}
\begin{center}
\begin{tabular}{cccccc}
\hline \hline \\
  I      &   II     & III            & IV            & V       & VI         \\   \hline
{\bf  13.04.94}&          &                &              &           &            \\
 23:13   &     I B  &          600   &        1.7   &    22.71  &   35.1     \\
 23:55   &    II B  &          600   &        1.45  &    22.67  &    35.5     \\
 00:25   &   III B  &          600   &        1.31  &    22.65  &    36.0     \\
 00:51   &    IV B  &          600   &        1.11  &    22.58  &    37.0     \\
 23:02   &     I V  &          600   &        1.47  &    21.83  &    35.7     \\
 23:33   &    II V  &          600   &        1.40  &    21.82  &    35.2     \\
 00:14   &   III V  &          600   &        1.14  &    21.79  &    35.8     \\
 00:39   &    IV V  &          600   &        1.08  &    21.72  &    36.4     \\
{\bf  2.06.94} &          &                &              &           &            \\
 19:12   &     I R  &          600   &        1.70  &           &    36.5     \\
 19:32   &    II R  &          600   &        1.63  &           &    37.1     \\
 19:51   &   III R  &          600   &        1.83  &           &    37.5     \\
 20:13   &    IV R  &          600   &        2.14  &           &    38.0     \\
{\bf  29.12.94}&          &                &              &           &            \\
 02:18   &     c B  &          500   &        1.59  &    22.70 &   37.9     \\
 02:45   &     c B  &          500   &        1.63  &    22.66  &    37.1      \\
 03:14   &     c B  &          500   &        1.56  &    22.11  &    36.3      \\
 02:27   &     c V  &          500   &        1.47  &    21.81  &    37.6      \\
 02:54   &     c V  &          500   &        1.59  &    21.71  &    36.7      \\
 03:23   &     c V  &          500   &        1.77  &    21.10  &    36.3      \\
 02:36   &     c R  &          400   &        1.41  &    20.54  &    37.3      \\
 03:03   &     c R  &          400   &        1.37  &    20.38  &    36.5      \\
 03:32   &     c R  &          400   &        1.61  &    19.52  &    35.9      \\
{\bf  24.04.95}&          &                &              &           &            \\
 22:03   &     a B  &          600   &        1.40  &    22.82  &    35.1     \\
 22:14   &     a V  &          600   &        1.29  &    22.03  &    35.1     \\
 22:36   &     a V  &          600   &        1.12  &    22.04  &    35.2     \\
 22:24   &     a R  &          600   &        1.24  &    21.02  &    35.1      \\
 22:47   &     a R  &          500   &        1.20  &    20.98  &    35.3      \\
 22:56   &     a R  &          500   &        1.26  &    20.98  &    35.4      \\
 23:21   &     b B  &          600   &        1.18  &    22.89  &    35.7      \\
 23:32   &     b V  &          600   &        1.28  &    22.10  &    36.0      \\
 23:52   &     b V  &          600   &        1.14  &    22.08  &    36.5      \\
 23:42   &     b R  &          600   &        1.25  &    20.88  &    36.2      \\
 00:04   &     b R  &          500   &        1.14  &    20.94  &    36.7      \\
 00:13   &     b R  &          500   &        1.05  &    20.81  &    37.0      \\
 01:14   &     c B  &          600   &        1.30  &    22.62  &    39.0     \\
 00:30   &     c V  &          600   &        1.20  &    21.50  &    38.8     \\
 00:55   &     c R  &          600   &        1.18  &    20.63  &    38.0     \\
 00:41   &     c R  &          600   &        1.14  &    20.60  &    38.5      \\ \hline
\end{tabular}
\end{center}
\end{table}

\section{Search strategy}

\begin{figure*}[t]
\centerline{
\vbox{\psfig{figure=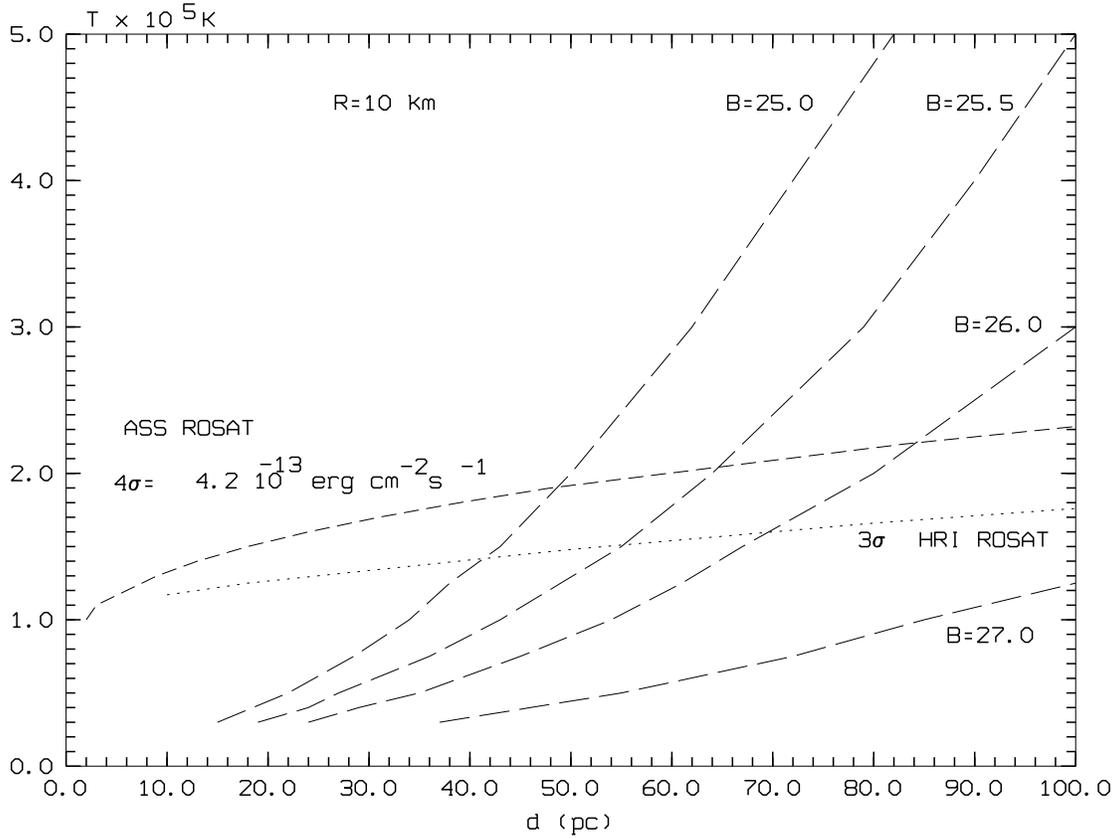,width=15cm,%
 bbllx=50pt,bblly=35pt,bburx=570pt,bbury=440pt,clip=}}
}
\par
\caption{
  Limits to  a black body surface
 temperature $T_{bb}$ and a distance of neutron star
type object with $R = 10\,  km$ as obtained from optical and ROSAT
observations are shown. The dashed line corresponds to a  black body flux
 $4.2\cdot 10^{-13}\,  erg\ cm^{-2}\ s^{-1}$ for the All--Sky Survey ROSAT
in the band $0.1 - 2.4\  keV$ or $4\sigma$ (Boer et al. 1993a, ROSAT, 1994b).
The dotted line corresponds  to $3\sigma$ of a limit with the $t_{exp}=13168\  s$
exposure for the ROSAT HRI camera for a black body point source ROSAT
(1994a). Long dashes denote optical limits of black body source in
the B band (see also the text).}

\end{figure*}

Let us assume that what we are looking for in  the  \g--ray  box
under investigation is indeed a single compact object of
neutron star type. For example, it could be an object  partially  or
totally consisting of quark matter, which has been  a hypothesis  so  far
(Haensel et al., 1986; Overgard \& Ostgaard, 1991).
 But here we try to formulate  a possible  observational
forecast  without  any  theoretical  model
describing \g--ray burst by itself. We do  not  assume  any
particular  burst model in advance. The only thing which can be  now
investigated from observations is the compact character  of  their
sources. Although till now the validity of absorption (cyclotron?)
features in redshifted annihilation lines observed in the spectra
of ``old'' \g--ray bursts is disputed and carefully tested,  the
observational fact of fast temporal flux variations in  bursts  is
not in  question (Atteia et al., 1987; Fishman  et al.,
  1994; Ryan et al., 1994).  It  means  that  up to  now  it   is
sufficiently reasonable  to  suppose  that  the  source  of  these mysterious
events is a ``compact  object''  with  probable  strong (not damped) magnetic
field.  \begin{figure*} \centerline{
\vbox{\psfig{figure=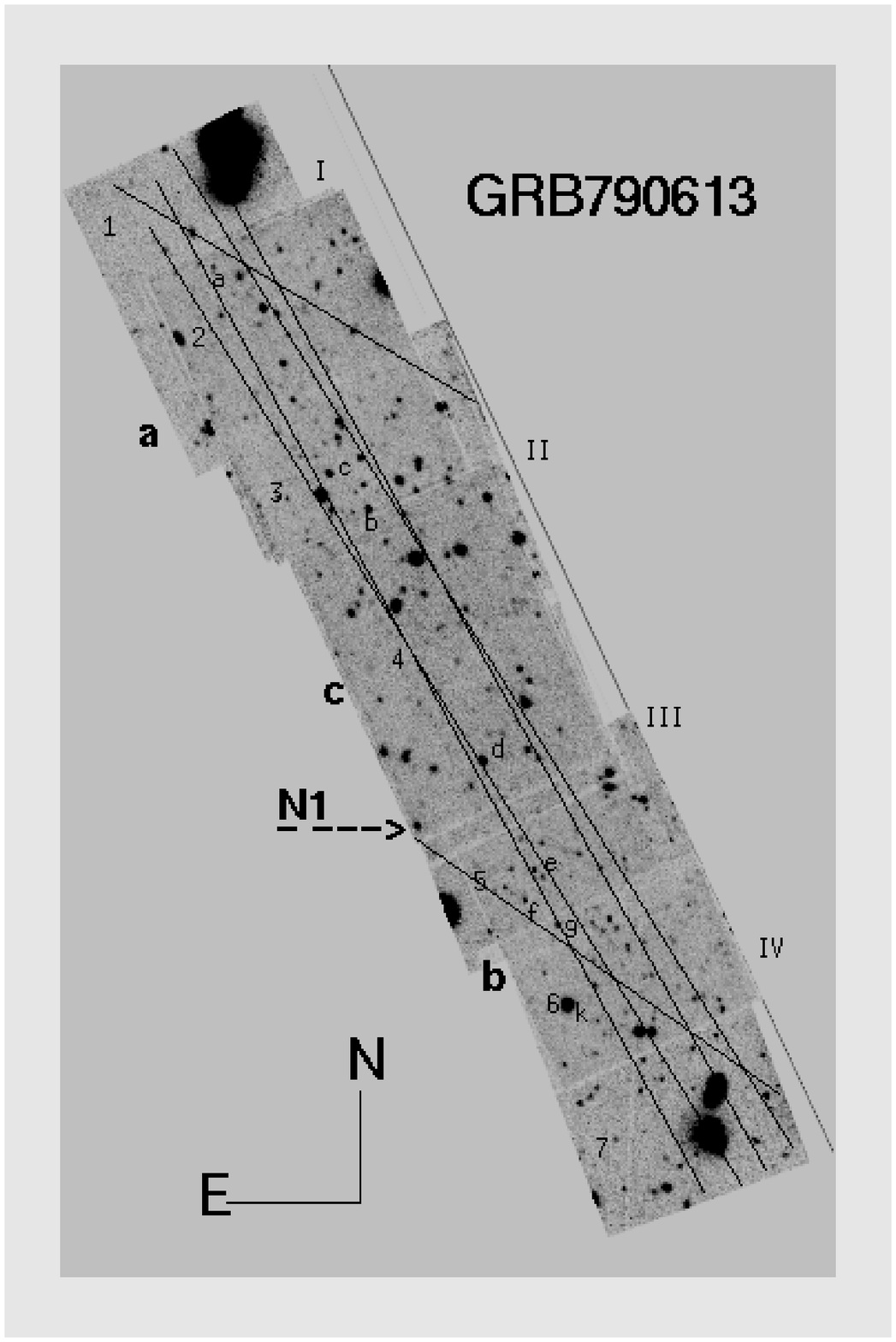,width=14.5cm,%
 bbllx=120pt,bblly=145pt,bburx=490pt,bbury=700pt,clip=}}
}
\caption{
 The CCD mosaic of the GRB\,790613 error box.
The error box was taken from Barat et al. (1984).
 Rome numbers and bold letters correspond to
numbers and letters in column III of the Journal of observations. Small letters correspond to
secondary photometric standard stars inside the CCD mosaic. Numbers correspond to
areas with homogeneous photometric limits.
Number 1 marks the star--like object with $B=23.03,\ B-V=0.03,\ V-R=0.1$.
}
\end{figure*}

Further we assume that potential sources of \g--ray  bursts  can
be among objects forming the hidden mass of the  Galaxy.  For  the
vicinity of the Sun, which we are basically interested here in, the
upper limit to the contribution of unknown objects to the  total
density of matter (or Oort's limit deduced from observed star
velocities  along   $z$   coordinate)   is   determined   and   is
approximately 38\% (Lacarrieu, 1971). Thus, we assume that the objects
sought may be rather old, in particular, their velocities can be close to
the velocities of the oldest pulsars. And the total expected number of such
objects in the optical and in the soft X--rays is bounded above by the
density of hidden mass,  which  gives  the  local  (in  the vicinity of the
Sun) numerical density of compact objects with the mass of $\rm
1.4\,M_{\sun}$ equal to $0.0357\,\rm  pc^{-3}$.

The total number of such objects can be bounded below by  the
number  of all neutron stars, which could be in the vicinity of the
Sun (Paczynski, 1990). But, apparently, old neutron stars (ONS) only can not
be  potential sources of \g--ray bursts, basically because  of  their
small number in the Galaxy,  which  demands  repeated  \g--ray bursts
at the average rate of events registered per  day  in  the BATSE experiment
(Fishman  et al., 1994;  Hartmann, 1994).  Proceeding  from  the
 population  $N_{tot}\sim 10^{9}$ of galactic neutron stars born in the
Galaxy disk with the velocity distribution observed for radio pulsars (as a
result,  part of them  will  leave  for  the  halo),  then  the spatial
density of our hypothetical objects must not be lower  than
$n_{ONS}\sim 0.001(N_{tot}/10^{9})\,\rm   pc^{-3}$,   based   on   the
estimates in (Paczynski, 1990) for a probable  average  density  of  old
neutron stars in the Galaxy plane near the Sun.

We will use $\rm R = 10\rm~  km\ {\rm and }\ M = 1.4~  M_{\sun}$ for the
radius  and the mass of such an object --- a hypothetical potential compact
source of \g--ray burst, since we assume a kind of  stable  object  or
stable configuration, close to  a  ``standard  neutron''  star  at
least by  their  gravitational  properties.  Below  all  estimates
concern these parameters namely, though, naturally, the radii and
masses of really observed neutron stars (pulsars)  are  determined
with a  dispersion  of  values.  To  estimate  expected
optical properties of these objects (in a quiet state --- before  or
after the burst), we should choose also a definite temperature  of
the thermal emission from the whole  surface  of  such  a  compact
object.

The limitation on the temperature $T_{bb}$ on the whole surface of
the compact object for  an  analogous  model  (a  ``cold''  single
neutron  star  as  a  potential  source  of  \g--ray  burst)  in
connection with deep X-ray EXOSAT observations of \g--ray  boxes
has been already used in the interpretation of optical observations  of
the supposed candidates near the  \g--ray  box  GRB  790325b (Hartmann  et
al., 1989; Motch et al., 1990).  There is already quite a number of
papers  limiting  the  neutron star surface temperature by observations not
only  with  the EINSTEIN and EXOSAT satellites (Pizzichini  et al., 1986;
Boer et al., 1988; Boer  et al., 1991), but also by later  ROSAT
data in the soft X--rays (Boer  et al., 1993b) (see also references in
Boer et al., 1993a; Greiner  et al., 1995). Here we use the results of
 ROSAT observations in the soft X--ray range (0.1 - 2.4 keV)
for  the same purpose --- to set an upper limit to  the  value
$T_{bb}$, to set later, using  our  own  optical
observations of  the GRB\,790613 \g--ray box, a  limit  on $T_{bb}$  and
a lower  limit  on the distance to the supposed potential candidates. Thus,
that is a kind of development of the  method of Pizzichini et al. (1986)
for single objects, which has  already become almost standard.

In  Fig.1  are  presented   corresponding   bounds   on   $T_{bb}$
temperature of emission from the surface of the compact object  with
$R=10\rm \ km$ as a function of distance to it.  The  source  of  black
body emission with $R=10\rm \ km$ can be detected in  the X--rays  or/and
in the optical if at a given temperature and distance it is found to the
left or/and above the corresponding curve in Fig.1.

For the all--sky survey in the energy range of $0.1-2.4\rm \ keV$  with
the help of the ROSAT satellite (ROSAT All-Sky Survey = ROSAT ASS)  we
will use the flux  $F_{x}\approx 4.2 \cdot 10^{-13}\,{\rm erg\cdot cm^{-
2}\cdot s^{-1}}$, averaged over the whole survey  (with an  average
exposure time of 500 s), which corresponds  approximately  to a
level of $4\sigma$. (5 X-ray sources found near  or  inside  small
\g--ray boxes were studied at this level of signal detection  in
ROSAT ASS (Boer  et al., 1993a)). If based on this flux value,  then  at
$T_{bb}  > 200000\rm~ K$, even for a spatial density (lower estimate) of
$0.001 \rm~ pc^{-3}$ in ROSAT ASS, there must be a considerable number (about
30 sources up to 20 pc) of  bright  ($\sim4\cdot10^{-12}\,{\rm  erg\cdot
cm^{-2}\cdot s^{-1}}$) X--ray sources,  which  could  be  identified
long ago if they existed indeed. So far there are  no  unidentified
{\it bright} X-ray sources (J.Greiner,  private  communication),
and from the absence of large number of coincidences of weak X--ray
sources with small \g--ray boxes (Greiner et al., 1995) it should be expected  that
even if such objects do exist in Nature, then their temperature is
lower than $1-2\cdot10^{5}$\,K, according to Fig.1.

At $T_{bb} \la 10^{5}\, \rm K$ even at a maximum  allowable  density  of  such
objects in the vicinity of the Sun ($0.0357\, \rm~ pc^{-3}$) in ROSAT ASS
a very small number should be expected of  corresponding  X--ray
sources  with a  typical   flux   of   $\ga 5\cdot10^{-13}\, {\rm   erg\cdot
cm^{-2}\cdot s^{-1}}$. At the same time, as is  seen  from  Fig.1,
corresponding compact objects with  temperatures  from  50000\, K
to 150000\, K will be seen as objects stronger than 25.5$^m$
in the B band, from 30 to 50\, pc respectively.

Consequently, in the vicinity of the Sun we are concerned with
the search in the optical for weak  blue  objects  with  colors:
$U-B\approx~-1.0,~B-V\leq~-0.3,\ {\rm and}~ V-R\approx -0.1$.  On  the
whole sky sphere  they  can  be  met  with  a  frequency  up  to
$\approx0.5/{\rm degree^{2}}$, if their temperatures are $\approx
10^{5}\rm~ K$.  At the same time in the all--sky  survey  in  the soft
X--ray  range such sources will be either weak (if
$T_{bb}\approx2\cdot10^{5}\rm~ K$)  or so weak (at $T_{bb}\approx10^{5}\rm
~K$) that they  can  be  detected only at observations with long exposures in
the same frequency band (see Fig.1).  The fact that              some source
was  found nevertheless in ROSAT ASS studying  40 small \g--ray boxes of a
 total area  of $\approx3.1{\rm \ degree^{2}}$ ( Greiner et al., 1995)
does not contradict the above  said and this implies that the temperature of
such objects is $\la 2\cdot10^{5}\rm \, K$.

Some contribution to the total bolometric luminosity can be  made
by accretion from interstellar medium onto  a  single  compact
object moving at a velocity of {\it v} (Bondi\, \& Hoyle, 1944).  In
the  computation  of the accretion rate $\dot{M}\sim n_{ISM}\cdot v^{-3}$ for
a given interstellar medium density $n_{ISM}$ (for the number of hydrogen
atom in a $\rm cm^{3}$) we will assume  that  the  velocity  of  our
hypothetical objects is close  to  the  velocity,  which  was  really
observed for radio pulsars, or,  at  least  is  close  to  spatial velocities
observed for the oldest of them (Lyne \& Lorimer, 1994).
Assuming that in the vicinity of  the  Sun  $n_{ISN}$  is  on
average    $\approx 0.07\, {\rm   H\ cm^{-3}}$,    then    at
$T_{bb}\approx10^{5}\,K$     the     corresponding      contribution
$(G\cdot M\cdot \dot{M}/R)$ to the total bolometric  luminosity  is
only several percent for $ v \ga 200{\rm ~km\cdot s^{-1}}$.

It is evident that because of their assumed low luminosity we  can
try to look for such objects only in the closest vicinity  of  the
Sun. It refers not only to the optical, but also to the soft  X--rays,  which
are strongly sensitive to inhomogeneities  in the  interstellar  medium
distribution (Paresce, 1984). So, keeping in mind the data  on
 interstellar medium close ($ \la 250$\, pc) to the Sun and also the
absorption not  only  in the soft X--rays, but also in the far ultra--violet
ranges (Welsh  et al., 1994), we  have  to do in the {\it optical} only  with
corresponding objects (observed both in the ultra--violet and in the soft
X--rays), which are situated very close to us. These may be only the
objects which  are  in  the bounds of the Local Bubble. Hence
our model estimates refer also only to the nearest space, where,  in
particular, the distances to the supposed sources are determined only by
their average spatial density ($\approx 0.001-0.036/{\rm pc^{3}}$), if it is
assumed  to  be approximately constant in a sphere  of   radius
100 pc.  Thus, first of all from the observational point of  view,  we  are
interested in the local situation with the  sources  of  \g--ray
bursts of the bright (short and powerful) events for which  the
law of -3/2  in  the  distribution  LogN/LogS  is  still  valid
(Kouveliotou et al., 1993; Fishman et al., 1994). Here we try  to  use
a  minimum  of  model  assumptions referring to ``Big Galactic models''
(Hartmann, 1994)  and,  so, we are dealing here only with the nearest
objects --- potential sources  of  \g--ray bursts.

GRB\,790613 is just  one  of  the  most  powerful  short  ($\approx 0.1$~s)
events which was simultaneously observed with the help  of  V-12
SIGNE, V-11 KONUS and PVO at photon energies exceeding 100 keV
($\approx 0.1{\rm \ photons\cdot cm^{-2}\cdot s^{-1}\cdot}$ keV), when the
greater part  of \g--ray burst energy was released at the  annihilation
emission frequency during a time of $\approx 0.1$\ s (Atteia et al.,
1987; Vrba et al.,  1995).  It could be
expected that a supposed compact object, which was connected with
this \g--ray burst, is one of such nearest objects, that has probably
undergone a kind  of  explosion  on its  surface.  In  next
Sections, with the use of the above mentioned, by  the example  of  GRB\,790613 event it will be shown how an observational problem of  search for
such ``cold'' compact objects can be stated under the  conditions,
when we have to conduct optical search for weak sources by studying
very big areas in the sky.

\section{Observations}

The  observational  data  were  obtained  with  the  CCD camera of $580\times
520$ pixels placed at the prime   focus of the 6 m telescope of SAO RAS.
 The CCD chip has  rectangular  pixels  of  $24\times18\  \mu$m, resulting in
an image scale of $0\farcs205\times 0\farcs154$  per  pixel.  We obtained
the data with Cousins {\it BVR} filters. Normal program  exposures were
600~s, 500~s or 400~s for the {\it BVR} filters. Longer exposures  are not
reasonable because of increasing number  of  space  particles, since about 30
particles on the average are  registered  during a 10 minutes' exposure. A
total of $\approx 5.7$ hours of open  shutter time was eventually obtained
for the observation of  the GRB\,790613 error box.  Atmospheric  conditions
were stable throughout the observational runs with a seeing of $1.1 - 1.7$
arcseconds.  The first observation of this box was carried out on 13/14 April,
1994  in the {\it B} and {\it V} filters (Sokolov et al., 1994), where
blue ``candidates'' were selected by the criterion  $B-V < 0$. An additional
observation in {\it R}  filter  was made on 2/3 June, 1994, but the CCD was
not in the standard mode, and we  used  these  data  only  as  a  photometric
reference  of relatively bright stars in the {\it R} filter over the whole
mosaic obtained later. A subsequent observation in the three ({\it BVR})
filters  of the central part of the GRB\,790613 error box, where several blue
($B - V < 0$) objects were found, was  carried  out  on  29/30  December,
1994, and for the whole field --- on 24/25 April, 1995.

The coordinates of the center of the studied field are $\alpha  (1950)
= 14^{h}12^{m}12^{s}.8;$ $\delta (1950) = +78^{o}54'13\farcs 8$. Unlike
other observers, we deliberately choose a symmetric \g--ray  box
obtained as a result of intersection of  three  stripes,  and  not
only its northern part, which  is  a  common  part  for  all  four triangulation
stripes, since the symmetry of the error box evidences in favor of
a correct burst localization. The total area of the symmetric  box
is $\approx 1.5\,{\rm \ arcmin^{2}}$.

The log--book is adduced in Table  1.  For the  GRB\,790613  field
there are: I -- the date and UT of observations, II  --
the area number and the filter, III -- the exposure time, IV -- the seeing, V
-- the night sky brightness per  one  square arcsec,    VI  -- the  zenith
distance.

The procedure of construction of the GRB\,790613  field  mosaic  we
obtained is shown in Fig.2.  The roman  numerals  and  bold  letters
correspond to the numbers and letters in column III of  the  log-book.
Small letters indicate seven secondary photometric standards.  The
numbers correspond to areas with homogeneous photometric limits (see Table
2).

\begin{figure*}
\centering{
\vbox{\psfig{figure=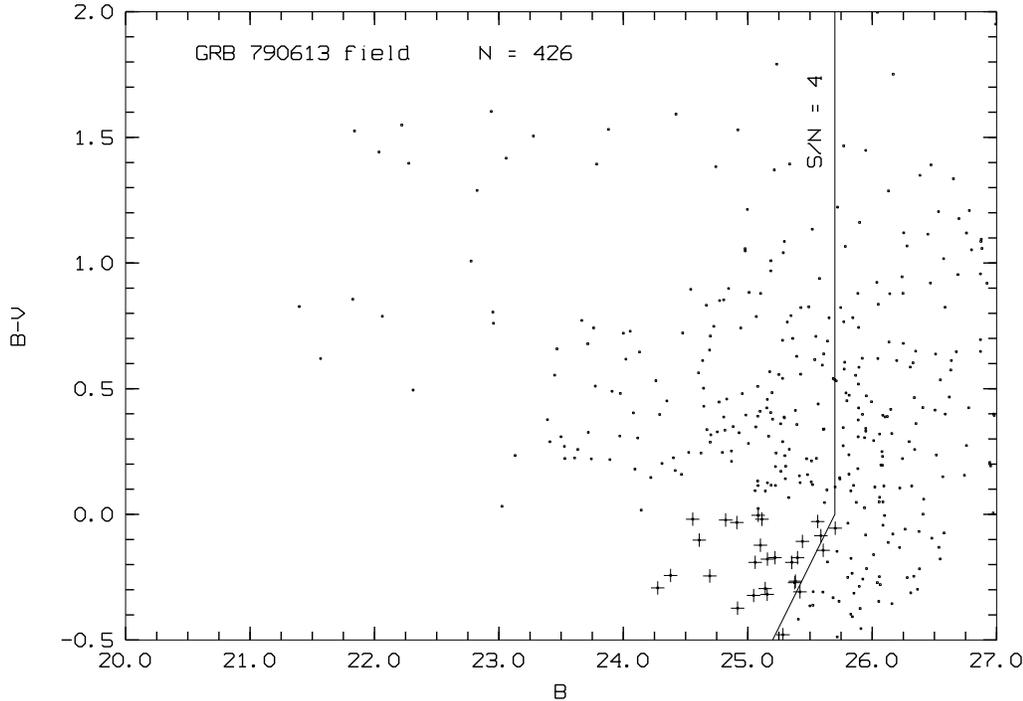,width=14cm,%
 bbllx=25pt,bblly=50pt,bburx=550pt,bbury=430pt,clip=}}  \par
}
\caption{
 $ B-V\ {\rm vs.}\ B$ is shown for all detected objects ($ N=426, B \leq 27$) of the GRB 790613 field .
 Crosses mark all objects with $B-V\ <\ 0$ and $S/N \geq 4$ . The line corresponds to the
``practical photometric limit'' ($S/N\approx 4)$.
}
\end{figure*}

\subsection{Data reduction and calibration}

Standard data processing involves bias removing, flat fielding and
removing of space particles  and  bad chip pixels. Bias-images
were obtained during the night  at  regular  intervals.  Flat-field
frames were obtained at twilight along with bias frames at the
end of each night of observation. By shifts and turns  all  images
were reduced to a unified  internal  frame  of  reference.  Cosmic
particles were selected by subtraction of  median  smoothing  over
several images from one image and were removed  individually.  The
whole processing was carried out with the  use  of  the  ESO-MIDAS
software package.
\begin{table}
\caption{
 The photometric limits.}
\begin{tabular} {cccc}
\hline \\
Number         &   B    &    V   &     R    \\
of area (Fig.2)&        &        &          \\  \hline
 1,7           &  24.7  &  24.7  &  ---     \\
 2.,3,5,6      &  25.7  &  25.7  &  25.6    \\
 4             &  26.2  &  25.8  &  25.6    \\   \hline
\end{tabular}
\end{table}

 The photometric standard star MS~1443+63\, B  was  observed  every
observational night before and after observation of the GRB\,790613
error box. We used the star MS~1443+63\, B to calibrate  secondary
standard stars inside the  CCD mosaic.  We  used  seven  secondary
standard stars for calibration of different frames for  one  night
and for different nights. The total number of  the objects registered
up to 27$^m$ in the {\it B} filter is 426.

\subsection{Photometry}

The images obtained and reduced to a unified  frame  of  reference
were summed up in each filter. Photometry  was  obtained  for  all
objects inside the whole CCD mosaic. But for further  analysis  we
used only the objects with a  $S/N  \geq  4$.  The
$S/N = 4$ corresponds to the  error of  0.25$^m$
and for an object with magnitude determined with such an error the
color error is 0.35.

The photometric procedure consisted of the following stages:
\begin{enumerate}
\item
 Since we were interested first of all  in  blue  ($ B-V   <0$)
objects, then by the images  obtained  in the {\it B}  filter  using  MIDAS
context  INVENTORY  and  smoothed  with  a  gaussian  we  obtained
coordinates for all detected objects of the whole mosaic.
\item By  the coordinates  obtained  at  the  first  stage  the   aperture
photometry was done in all three {\it BVR} filters.
\item The obtained values were corrected for the  finite  radius  of
the aperture and atmospheric extinction at the given zenith  distance,
and were reduced to the standard Cousins system.
\end{enumerate}
The results of the photometry are shown in Figs. 3,4,5.  The crosses  denote
 31 objects in Figs. 3,4 with $B-V \ <\ 0$ and  $S/N  >  4$.  Photometric
limits (defined as $ S/N=4$) for the  CCD mosaic are given in Table 2.

The  $S/N$ was calculated for every object by the formula:
$$ \frac{S}{N} = \frac{F_{star}}{\sqrt{F_{star}+F_{sky}}}, $$  where
$F_{star}$ is  the  flux  from  a  star  in  the  given  aperture,
$F_{sky}$ is the sky background in the same aperture.

\begin{figure*}[t]
\centering{
\vbox{\psfig{figure=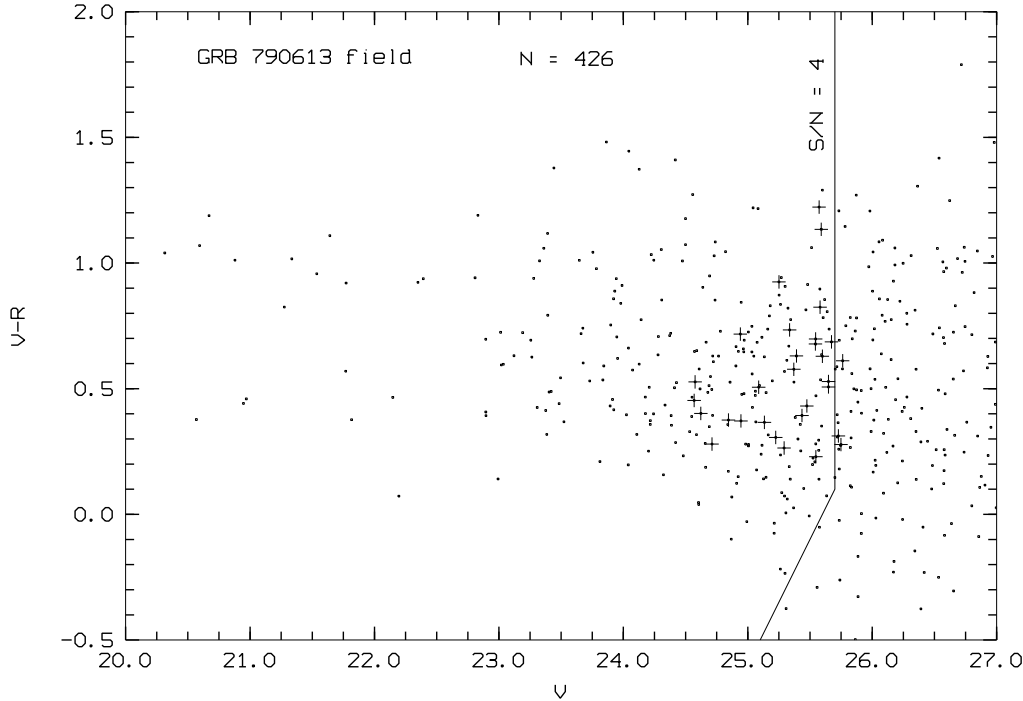,width=14cm,%
 bbllx=25pt,bblly=50pt,bburx=550pt,bbury=430pt,clip=}}\par
}
\caption{
 $ V-R\ {\rm vs.}\ V$ is shown for all detected objects ($ N=426, B \leq 27$)
of the GRB 790613 field . Crosses mark 31 objects with $B-V\ <\ 0$ and $S/N >  4$.
}
\end{figure*}

Photometric errors were determined as follows:
\begin{enumerate} \item If we had one measurement only,  then  the  error
was  calculated using Poisson distribution of accumulated quanta:  $$
\sigma_{mag} = \sqrt{ \sigma_{star}^{2} +  \sigma_{sky}^{2}},$$ where
$$\sigma_{star} = \sqrt{F_{star}},\, \sigma_{sky} = \sqrt{F_{sky}}.$$ \item
If we have several measurements, then in addition to 1)  $$ \sigma_{mag}
  = \sqrt{\frac{\sum_{i}^{n} (mag_{sr} - mag_{i})^{2}}{n-1}}. $$ And
correspondingly, the color errors are obtained as follows: $$ \sigma_{(B-V)}
    = \sqrt{ \sigma_{B}^{2} +  \sigma_{V}^{2}},\,  \sigma_{(V-R)} = \sqrt{
    \sigma_{V}^{2} +  \sigma_{R}^{2}}.$$  The color errors  for  several
objects  with  color  indices  closest to those of probable black--body
objects are shown in Fig.5.
 \end{enumerate}
\begin{figure*}[t]
\vbox{\psfig{figure=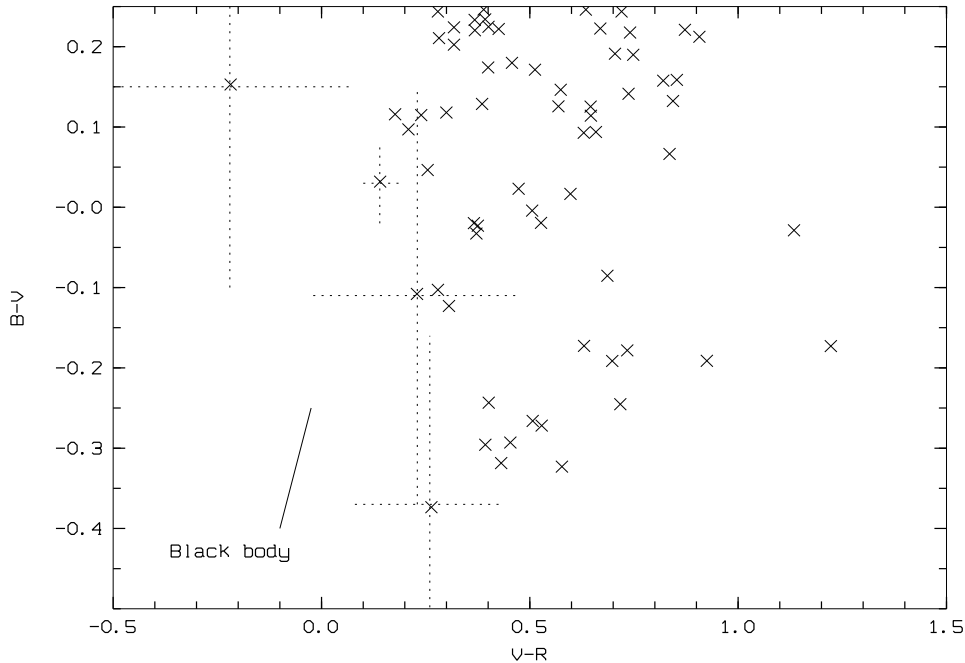,width=15cm,%
 bbllx=20pt,bblly=30pt,bburx=650pt,bbury=420pt,clip=}}
\par
\caption{
 $(B-V)/(V-R)$ diagram for relatively blue ($B-V < 0.25$) objects
 with the ratio $S/N > 4$.
A place is marked by a solid line
 where black body
objects with $100000\ {\rm K}\ >\ T_{bb}\ >\ 30000\ {\rm K}$
 could be situated.
For several objects being near this place
corresponding rms for their colors are indicated.
}
\end{figure*}

\subsection{The search for blue stellar--like objects}

The search for a neutron  star  type  object  with  a  black--body
surface  temperature  $T_{bb}\approx  10^{5}\rm \ K$,  as  a  possible
potential source of a $\gamma$-ray burst, in the case  $\gamma$-ray bursters
are of  local Galactic origin,  was the main
purpose of our observations.  These sources must be observed as faint
($ V>23^{m}$) blue stellar--like objects of $ B-V \approx -0.37\ {\rm and}\
\ V-R \la -0.1$ and may be soft X-ray sources.

That is why our first step in selection of probable ``candidates''
was selection of  objects  with  color  indices  $B-V < 0$  (see
Fig.3). Observations in the R filter (the second approximation)  allow
us to select among such blue ``candidates'' only the  ones, which
must be below the line $V-R=0$. In Fig.4 it is well seen  how  the
objects with $B-V \leq 0$ (crosses) ``scatter'' over the diagram  $(V-
R)/V$. A subsequent test of blue in $B-V$ objects with the help of
the criterion  $V-R<0$ reduces sharply large number  of  such
``candidates'' with the expected black body (``stellar'') distribution
over the $B, V, R$ fluxes even without a careful analysis of their  profiles
for the purpose of revealing stellar likeness or   extended  character
of an object.

As is seen from Figs. 3,4 and Table 2, the limits, from  which  the
level of  $S/N$ is greater than 4 for each  of the {\it B,  V,  R}
filters  in  the  average  for  the  whole   mosaic,   are   equal
correspondingly to $B=25.7,\ V=25.7$ (and $R=25.6$). Thus,  from  the
aforesaid it might be assumed that almost all obviously  blue  in
$B-V$ objects studied up to $B\approx25.2$ (see Fig.3)  are  too
red in color index $V - R$ to  attribute  them  confidently  (with
$S/N > 4$) to objects with the black--body  distribution  in the {\it B,  V,
R} fluxes. That is  no blue optical counterparts by $ B-V$ \rm and  $
V-R$ (for $ S/N > 4$) were found.

The ROSAT HRI observation of the GRB\,790613 field with  a 13168  s
exposure did not detect any objects in this error box  and  showed
that the black body temperature must be less than 150000  K  for
the objects of the neutron star type at a distance less than 100 pc (ROSAT
 X--ray Images, 1994a; ROSAT Guest Observer
 Program, 1994b). Our observation with ROSAT data can be interpreted as the
absence of a single compact object with $ R  =  10\ \rm km$  and  $
T_{bb}\approx 10^{5}\rm\ K$ up to a distance  of  $  \approx  40\ \rm  pc$
(Fig.1)  and accordingly, the energy of GRB 790613 event is higher
than $ (4 - 9.5) \cdot10^{35}\ \rm erg$. Nevertheless, if the value
$B\approx25.2$ is assumed to be the photometric limit, at which the  blue
object with $B-V\approx -0.3, V-R \ga -0.1$ can yet be  classified  (by  two
color indices at once!)  by our  data  (bearing  in  mind  an  object
analogous in brightness and color to Geminga  optical  candidate).
Then from the X-ray $3\sigma$-limit for deep observations of the  same
box with HRI ROSAT  (ROSAT X-ray Images, 1994a;
 ROSAT Guest Observer Program, 1994b) it is seen (Fig.1) that in this case  a
single object, which could be connected with this brightness, can be
situated very close to us, if its temperature  is  $T_{bb}  \la
10^{5}\ \rm K$.

\subsection{Estimations of proper motions}

Close objects (of NS type)  can  have  considerable  proper
motions. For example, at a velocity of $\approx\,200\,{\rm  km/s}$ and a
distance of 100 pc the proper motion up to 0.4 arcsec/year can  be
expected from such an object. But if the  object  has a  non--thermal
spectrum, it may be photometrically lost among the great number of
other objects. That is why the proper  motion  measurement  is  of
special importance as an opportunity to select {\it all} close
objects with high spatial velocities. The proper motion measurement
procedure was as follows.
\begin{figure*}[t]
\centering{
\vbox{\psfig{figure=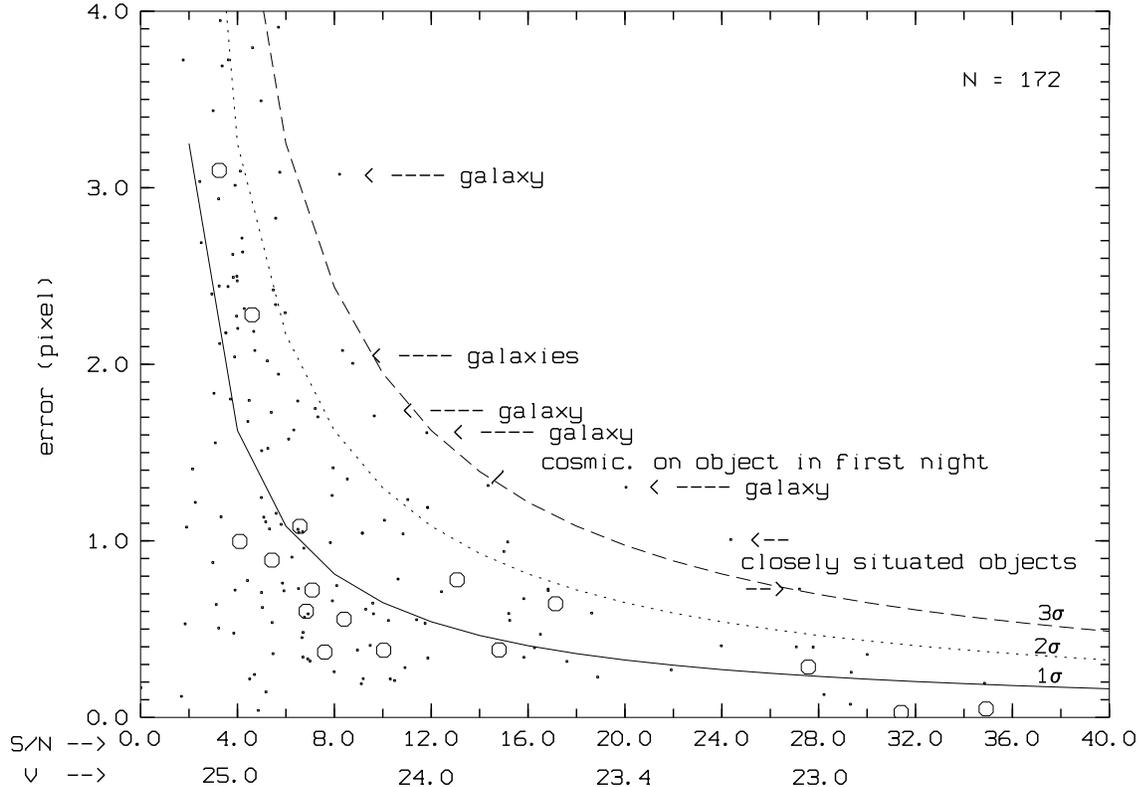,width=15.5cm,%
 bbllx=10pt,bblly=30pt,bburx=570pt,bbury=430pt,clip=}}  \par
}
\caption{
 The objects position error (point) was measured as a
difference of positions of the Gauss sum of {\it B} and {\it V }
images for 13/14 April 1994 and 24/25 April 1995. Circles are
 the  measurements of an object positions during one night.
Lines correspond to $1\sigma , \ 2\sigma ,\ 3\sigma$ of an
 average position error, $1 {\rm pixel} = 0\farcs205$. The axis with {\it S/N} corresponds to
{\it S/N} for Gauss sum of {\it B} and {\it V} images for April 1994.
}
\end{figure*}

 All images were reduced to one frame  of
reference with the help of bright stellar--like objects with normal
stellar color indices. Since these objects lie  in  the  range  of
20-21$^m$, they are the most probable stars at  the  distance
of several kiloparsecs and, correspondingly,  are  good  reference
points for construction of motionless frame  of  reference.  Since
all objects, which are  the  nearest  to  us,  must  have  stellar
profiles, to determine the position of all  objects  we  used  the
approximation of  their  profiles  by  the  Gauss  function,  and,
correspondingly, defined the position of the object as  the  Gauss
function center of mass. After making such  measurements  for
two epochs: April 1994 and April 1995, we  estimated  the  proper
motion  of  objects  per  year  as  the  difference  between   the
coordinates  for  the  two  epochs.  For  galactic  objects  this
difference would be just the proper motion, for extragalactic ones
it corresponds to the accuracy of the object profile approximation
by the Gauss function. To determine  the  standard  error  of  the
object position, analogous measurements were made for the same
objects in different images obtained during one night. The results
of our measurements are presented in Fig.6.

    The proper motion  of objects with  $V\leq  25$  was
measured. Down to $ V \approx 25$  objects with the proper  motion
 larger than $ 3\sigma$ of average position were not  found.
Objects with cosmic particles or closely spaced objects have a
position error between $ 2\sigma\ {\rm and}\ 3\sigma$  of  the  average
position error.

\begin{table}[t]
\caption{
The photometry of bright galaxies of the group.}
\begin{tabular} {cccccc}
\hline \\
 Number & B    &    V   &    R   &  B-V  &  V-R         \\
of galaxy&   &        &       &        &       \\
(Fig.8) &    &        &       &       &      \\ \hline
 1. &  21.39 &  20.42 &  19.65 &  0.97 &   0.76        \\  
 2. &  21.44 &  20.35 &  20.03 &  1.08 &   0.32         \\ 
 3. &  20.84 &  19.86 &  19.54 &  0.97 &   0.32         \\ 
 4. &  22.03 &  21.00 &  20.53 &  1.02 &   0.47         \\ 
 5. &  22.13 &  20.82 &  20.27 &  1.31 &   0.55         \\ 
 6. &  22.60 &  21.64 &  21.23 &  0.96 &   0.40         \\ \hline
\end{tabular}
\end{table}

\subsection{Some interesting objects}

Some interesting objects were found near and inside the error box.
The first (Number 1 in Fig.2) of them  is  a  stellar-like  object
with $B=23.03$, $B-V=0.05$ and $V-R=0.1$. It is  situated  outside
the error box at a distance of about $ 30^{\prime \prime}$ from  its
boundary.  The object has color indices close  to  those of black  body  with
the temperature $ T_{bb}\approx 17000\ \rm K$. Assuming  that  this
object is the object of  neutron star type, then  it  might
be situated at a distance less than  10 pc  from  the  Sun  and,
accordingly, it has to have a noticeable  proper  motion  ($  >  1\
{\rm arcsec/year}$).
\begin{figure*}[t]
\centering{
\vbox{\psfig{figure=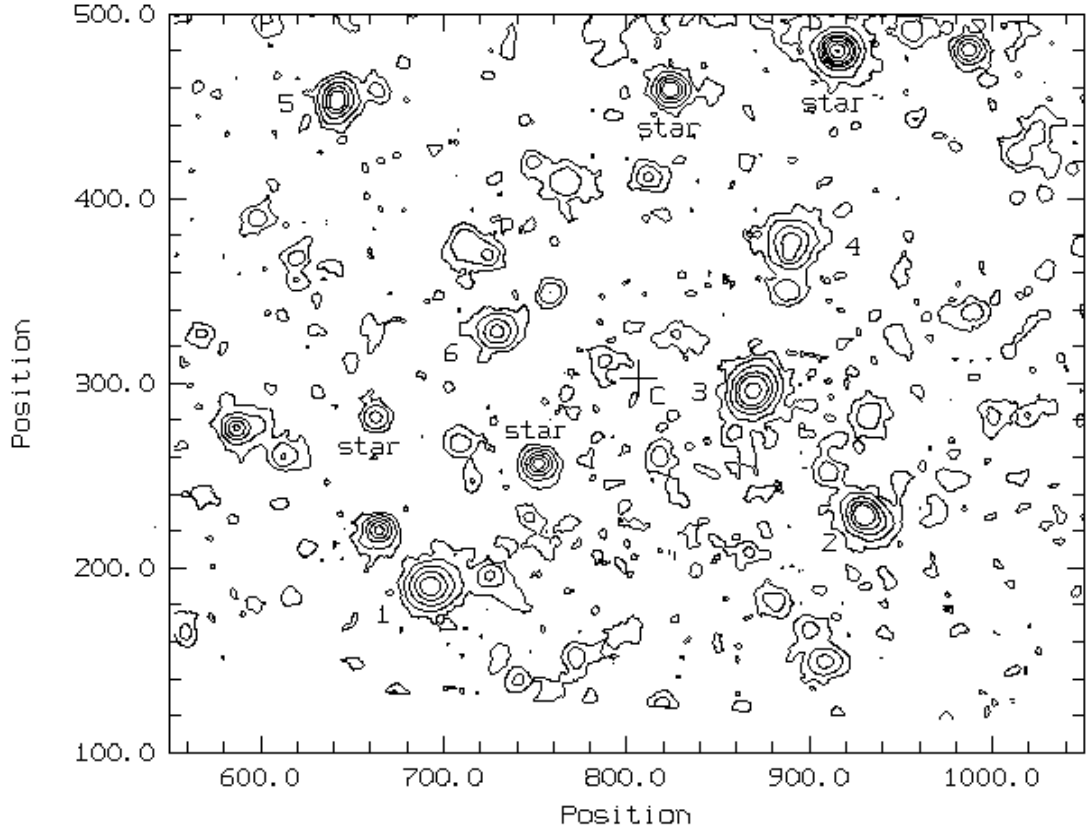,width=16cm,%
 bbllx=-100pt,bblly=190pt,bburx=570pt,bbury=655pt,clip=}}  \par
}
\caption{
  The isophote map of the group of galaxies is shown. Numbers correspond to
numbers in the Table 3  with the photometric data for these galaxies. The cross marks the
center of mass for this group.
}
\end{figure*}
 But the proper motion of this object is less than $
0\farcs1$ per year. It may be a QSO or a halo object:  a  subdwarf  or  a
white dwarf at about 3 kpc distance.

A group of galaxies was found in a projection on the northern part
of the error box using the galaxy  counts.  Fig.9
displays the excess of bright galaxies when counting  them in  two
adjacent frames of our mosaic (II and III of  our  log-book and see
also Fig.2).  The size of the group of galaxies is about $
120\arcsec$.  It is obtained on the basis of counting the number of
galaxies up to $\approx 24.5^m$  in the  stripes  of
$80\arcsec \times 10\arcsec$ in size along  the whole  mosaic (Fig.8).
Assuming the masses of galaxies to be proportional to their  magnitudes,  the
determined coordinates of the center of mass for the group are  ${  {
\alpha(1950)  = 14^{h}12^{m}16^{s}7}}$,   ${ \delta(1950)   =
78^{o}55^{\prime}02\farcs 0}$.  The center of mass is denoted by the  cross
in Fig.7. The redshift of the group  $ z \approx 0.3$ is estimated from
magnitudes and colors of galaxies with the use of  K-correction taken from
the paper (Frei \& Gunn, 1994). The numbers in Table 3 correspond
to the numbers in Fig.7.

\begin{figure}
\centering{
\vbox{\psfig{figure=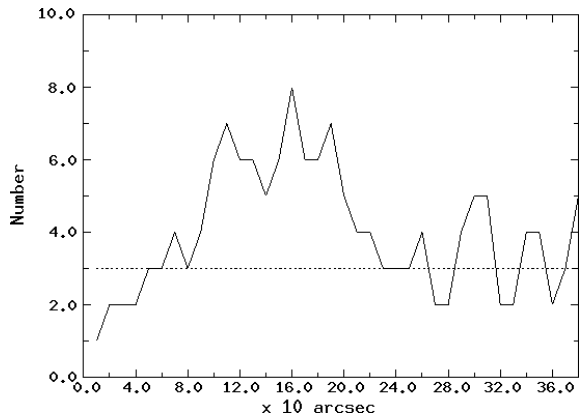,width=8cm,%
 bbllx=40pt,bblly=30pt,bburx=490pt,bbury=350pt,clip=}}  \par
}
\caption{
 The size of the group is $120''$.
The size was obtained using counts of galaxies up to $V\approx24.5$ in  $80\times10$ arsecond strips along the CCD mosaic.
}
\end{figure}

\begin{figure}
\centering{
\vbox{\psfig{figure=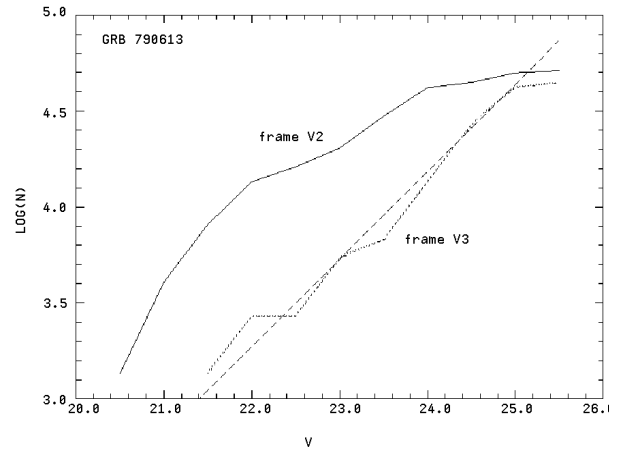,width=8cm,%
 bbllx=77pt,bblly=53pt,bburx=459pt,bbury=350pt,clip= }}  \par
}
\caption{
 The counts of galaxies in two neighbouring frames. The frame N2 corresponds to II {\it V} in the
Journal of observation. The frame N3 corresponds to III {\it V}. Log(N) is logarithm of number of galaxies per
a square degree and per a magnitude.
}
\end{figure}

\section{Summary and discussion}

From the mentioned in Section 3.3 it follows  that  today's
X-ray and optical observations do not exclude at all  a  cooled  compact
object as a candidate for the sources of \g--ray  burst  in  this
concrete case of the \g--ray box, which has already been well  studied
in the X--rays (see also in  Greiner et al., 1995). In general, based on
the average  limit of ROSAT ASS (dashed line in Fig.1), all such sources or
at least the closest of them could be located at a distances of
$\approx 50\ {\rm pc}$ or more  by  varying  the  spatial  density  of
corresponding hypothetical  cold  $(T_{bb}\leq  1.5  \cdot  10^{5}\ \rm K)$
compact objects from a maximum density  of  $0.0357/{\rm pc^{3}}$  (bearing
in mind the objects with the mass of  $1.4\,M_{\sun}$  as  unknown  objects
of (Lacarrieu, 1971) and up to a minimum spatial density close to the
density  of old neutron stars in the vicinity of the Sun.

It means that we can construct any ``global'' galactic models  for
\g--ray bursters, beginning with the well studied situation in the
vicinity of the Sun. And from Fig.1 it is well seen that if it  is
difficult to observe such single objects in the soft X--rays (because of
absorption also), then nevertheless at big telescopes  they  could
be detected in optics in small \g--ray boxes ($\approx 1'\times1'$)
up to a distances of $\sim100$ pc (up to $m\sim 27^m$).

The emission of these cold compact objects can be weakened in the soft
X-rays because of the fact that for the X--rays  the  walls  of  so  called
``Local Bubble'', which are  situated  close  to  the  Sun,  become
opaque (Paresce, 1984). (See the calculation of soft  X-ray  absorption  for
the thermal emission of single neutron stars  in Helfand  et al. (1980),
analogous calculation for the band of $0.1-2.4$\, keV is in Boer,  et al.,
(1993a).)  For  the  same reason there must be few
such objects in EUVE and ROSAT  WFC  sky surveys (Welsh et al., 1994),
and the objects which can still  be  seen  in  the far ultra--violet
(the closest ones) may be potential sources,  which have not yet
manifested themselves as the sources of \g--ray  bursts --- the sources
``in a quiet state''.

To emphasize here the  fact  that   study  of  blue  ($B-V<0$)
objects with $B\approx 25.5$ in  the  whole  field  is  not  a  limiting
problem for  the  6 m  telescope  at  all,  we  refer  to
analogous observations in February 1987 of the Geminga object  (Halpern  \&
Tytler, 1988) at the 5 m telescope. The corresponding exposures in the g and
r bands were 7200 s and 5500 s, which allowed  only a  rough
 color estimate ($g - r = -0.3$) to be obtained for the expected optical
candidate with  $g  = 25.1$, i.e.  this was  limiting problem
indeed.  Even in the  case of observation at NTT of an optical candidate for
PSR 0656+14 in January 1991 (Caraveo  et al., 1994), when an object with
$V\approx 25$ was found  at the level of $3\sigma$,  the  total  exposure
in  the {\it V} filter was 4200 s.  In the {\it B, V} and R frames obtained
at the 6 m telescope (during an exposure time of $\approx 25$\ min in
each filter, as was mentioned  above) all these objects were not only
detected at the level higher  than $3\sigma$ but their colors were
also measurable.  In particular, the object analogous to the optical
candidate for Geminga should be registered in {\it  B} and {\it V} with a
good {\it S/N.}

Certainly, in principle, the situation is possible when a probable
source of the $\gamma$-ray burst is  on  the same  line  of  sight  as  a
brighter object. The probability of such a projection of  the  GRB\,790613
event source on one of the relatively bright objects inside the
mosaic is $\approx$ 6\%. This is the ratio of the area  covered  by  all
objects to the total area of the mosaic for the  GRB\,790613  error
box. Thus, the observation of the  GRB\,790613  \g--ray  box  in
three filters allows us to hope for a 94\% ``cleaning''  of  the  whole
investigated field for such objects as optical candidates for  the
identification of the Geminga object (Halpern \& Tytler, 1988) or PSR~0656+14
(Caraveo et al., 1994).

Thus, the results of the study of the localization region  of  the
GRB\,790613 event up to the magnitudes  $B=25.7, V=25.7$ and $ R=25.6$
permits the following conclusions to be drawn:
\begin{enumerate}
\item
The absence of blue (by $B-V$ and $V-R$ color indices)  objects  up
to $B=25.2$ near the box center can be interpreted as the  absence
of a compact object with a ``naked'' surface  (without  any  dense
envelope in the optical) hotter then $\approx 150 000 \ {\rm K}$, at least
up to $\approx 40\ {\rm pc}$ in this direction. So,  optical  observations
add new details to the  estimate  of   temperature  of the supposed
single compact objects in the vicinity of the Sun,  which  can  be made in
interpretation of results of  the  search  for  point-like X-ray sources in
the range of $0.1-2.4$ keV.

\item The observational data available do not reject  the  existence
of ``cold'' ($T_{bb}\la10^{5}\ {\rm K}$) compact  objects  up to $\approx
2\cdot 10^{4}$, some of
which could be  related  with  the  closest  sources  of \g--ray
bursts in the vicinity of the Sun (in a sphere of  radius   $\la
50\ {\rm pc}$).  Accordingly,  lower  values  of the supposed spatial density
($\la 0.001/{\rm pc^{3}}$) at a given  observed  average frequency of \g--ray
events ``remove'' the closest source of the burst to a distance greater
than 50 pc. For a  given rate  of \g--ray bursts,
$<0.8{\rm /day}$  (Fishman et al.,
1994) it turns out  automatically that  old  neutron stars born in
accordance with the ``standard'' scenario (Paczynski, 1990)  can not be
sources of \g--ray events at least near the Sun).
\end{enumerate}
Since here we mean  only  the  most  powerful  and  the
closest to us sources of \g--ray  bursts  and,  correspondingly,
the closest compact objects probably related  to  them,  then
the global spatial distribution of sources  is open to question
and it becomes  essential  for weak sources   (Fishman et al., 1994).  The
question is also open of  recurrent  events,   short  and  long
flashes and, in general, the mechanism  of   \g--ray
burst remains incomprehensible.

The work was carried out with the support of  ESO  C\&EE  Programme
(grant A-02-023), and also with the support of Russian  Foundation
for Basic Research (grant 94-02-04871a).

\end{document}